\PassOptionsToPackage{table}{xcolor}

\documentclass[sn-mathphys,Numbered]{sn-jnl}

\usepackage{graphicx}%
\usepackage{amsmath,amssymb,amsfonts}%
\usepackage{amsthm}%
\usepackage{mathrsfs}%
\usepackage[title]{appendix}%
\usepackage{textcomp}%
\usepackage{manyfoot}%
\usepackage{listings}%
\usepackage[T1]{fontenc}
\usepackage[labelfont=bf]{caption}
\usepackage[labelformat=simple,labelfont=bf]{subcaption}

\usepackage{booktabs,multirow}
\usepackage{array}
\usepackage{pifont}
\usepackage{bbding}
\usepackage{pdfpages}
\usepackage{enumitem}
\usepackage{newtxmath}
\usepackage{wrapfig}
\usepackage{multicol}
\usepackage[ruled]{algorithm2e}

\SetCommentSty{mycommfont}
\usepackage{varwidth}
\usepackage[table]{xcolor}

\theoremstyle{thmstyleone}%
\theoremstyle{thmstyletwo}%

\theoremstyle{thmstylethree}%

\newcommand{\cm}[1]{{#1}}
\newcommand{\cg}[1]{{#1}}

\makeatletter
\newcommand*{\rom}[1]{\expandafter\@slowromancap\romannumeral #1@}
\makeatother

\begin{document}


\title[FastEndoNeRF-Sim]{Efficient EndoNeRF Reconstruction and Its Application for Data-driven Surgical Simulation}


\author[1]{\fnm{Yuehao} \sur{Wang}}
\author[1]{\fnm{Bingchen} \sur{Gong}}
\author[1]{\fnm{Yonghao} \sur{Long}}
\author[2]{\fnm{Siu Hin} \sur{Fan}}
\author*[1]{\fnm{Qi} \sur{Dou}}

\affil[1]{\orgdiv{Department of Computer Science and Engineering}, \orgname{The Chinese University of Hong Kong}, \orgaddress{\city{HKSAR}, \country{China}}}

\affil[2]{\orgdiv{Department of Biomedical Engineering}, \orgname{The Chinese University of Hong Kong}, \orgaddress{\city{HKSAR}, \country{China}}}


\abstract{
\textbf{Purpose} The healthcare industry has a growing need for realistic modeling and efficient simulation of surgical scenes. \cm{With effective models of deformable surgical scenes, clinicians are able to conduct surgical planning and surgery training on scenarios close to real-world cases.} However, a significant challenge in achieving such a goal is the scarcity of high-quality soft tissue models with accurate shapes and textures. To address this gap, we present a data-driven framework that leverages emerging neural radiance field technology to enable high-quality surgical reconstruction and explore its application for surgical simulations.

\textbf{Method} We first focus on developing a fast NeRF-based surgical scene 3D reconstruction approach that achieves state-of-the-art performance. This method can significantly outperform traditional 3D reconstruction methods, which have failed to capture large deformations and produce fine-grained shapes and textures. We then propose an automated creation pipeline of interactive surgical simulation environments through a closed mesh extraction algorithm.

\textbf{Results}  Our experiments have validated the superior performance and efficiency of our proposed approach in surgical scene 3D reconstruction. We further utilize our reconstructed soft tissues to conduct FEM and MPM simulations, showcasing the practical application of our method in data-driven surgical simulations.

\textbf{Conclusion} We have proposed a novel NeRF-based reconstruction framework with an emphasis on simulation purposes. Our reconstruction framework facilitates the efficient creation of high-quality surgical soft tissue 3D models.
With multiple soft tissue simulations demonstrated, we show that our work has the potential to benefit downstream clinical tasks, such as surgical education.

}




\maketitle

\section{Introduction}\label{sec1}


The development of realistic robotic surgery scenes is important for VR-based surgical training.
The conventional method for creating these surgery scenes involves manual creation of soft tissue models with in-vivo textures by skilled artists. However, this approach is highly time-consuming and restricts the level of detail and variety achievable in surgical simulation. To overcome these limitations, we propose an automated approach to reconstruct interactive surgical environments using captured real data.

Surgical reconstruction \cite{liu2020reconstructing,chen2020improved,recasens2021endo,wei2021laparoscopic,wei2021stereo,long2021dssr,wang2022neural}, as an emerging task, aims to recover the 3D shapes and appearance of soft tissues from in-vivo surgery videos. As pointed out by previous literature \cite{long2021dssr,wang2022neural}, surgical reconstruction is cursed with three typical challenges over natural scene reconstruction: 1) Soft tissues will undergo large and drastic deformations. Many surgical operations, e.g., cutting and tearing, can even damage the topologies of soft tissues. 2) Surgical tools usually appear on the surgery videos and partially occlude underlying soft tissues from observation. 3) Endoscopic surgery videos are captured in confined in-vivo spaces, resulting in limited multi-view geometric clues of the 3D shapes.
Our recent work EndoNeRF \cite{wang2022neural} exploits the strong capacity of NeRF \cite{mildenhall2020nerf} for scene representations and incorporates tailored modules for handling tool occlusion and single-viewpoint input, achieving significant improvements in surgical reconstruction, particularly for scenes with large deformations.
However, EndoNeRF encounters new practical challenges when constructing surgical simulation environments.
First, the process of reconstructing a surgical scene from endoscopic videos using EndoNeRF is inefficient, requiring over 10 hours for per-scene optimization. 
Second, the optimized geometry of EndoNeRF is represented in a purely implicit field, i.e., the whole scene is encoded by network parameters.
However, many physically-based methods in soft-body simulation \cite{muller2007position,sifakis2012fem,qian2017essential,qian2016energized} require explicit geometry model, e.g., meshes, particles, or tetrahedrons, rather than implicit fields.
It is also worth noting that the realistic interaction of soft tissues is reliant on the underlying content beneath the tissue surface. While the geometry in the EndoNeRF only represents the surfaces of soft tissues. Hence, apart from surface reconstruction, another significant challenge lies in recovering topologically closed counterparts of soft tissues for simulation purposes.

To fill this gap, this work is the first attempt to create surgical simulation environments with soft tissue surfaces automatically reconstructed from endoscopic surgery videos. Technically, we propose a novel framework for dynamic surgical reconstruction, which can yield realistic and simulator-friendly counterparts of the soft tissues in the input robotic surgery videos. We summarize our main contributions as follows:
\begin{itemize}
    \item We adopt a novel voxel grids-based scene representation for faster dynamic surgical scene reconstruction.
    \item We build a pipeline for converting radiance fields into a closed mesh, which enables physically-based simulation of the reconstructed surgical scenes.
    \item We exhibit multiple robotic surgery simulations with our reconstructed soft tissues on multiple simulation engines, including Taichi MPM \cite{hu2019taichi,hu2018moving} and NVIDIA Isaac Sim \cite{liang2018gpu}.
\end{itemize}
\vspace{0.1in}
This work builds upon a preliminary version presented at MICCAI 2022 \cite{wang2022neural}. In this paper, we have made significant revisions and extensions to the original conference version. The major improvements include:
\begin{itemize}
    \item[--] We designed a new deformable scene representation with grid-based radiance fields and 4D tensor-decomposed motion fields for faster training convergence.
    \item[--] We proposed a novel pipeline for extracting closed meshes from radiance fields, in order to generate simulatable soft tissues.
    \item[--] We conduct multiple surgical scene simulations with our reconstructed soft tissues.
\end{itemize}
Our code is available at \href{https://github.com/med-air/EndoNeRF}{https://github.com/med-air/EndoNeRF}.

\section{Method}

\begin{figure}[t!]
    \centering
    \includegraphics[width=1.0\textwidth]{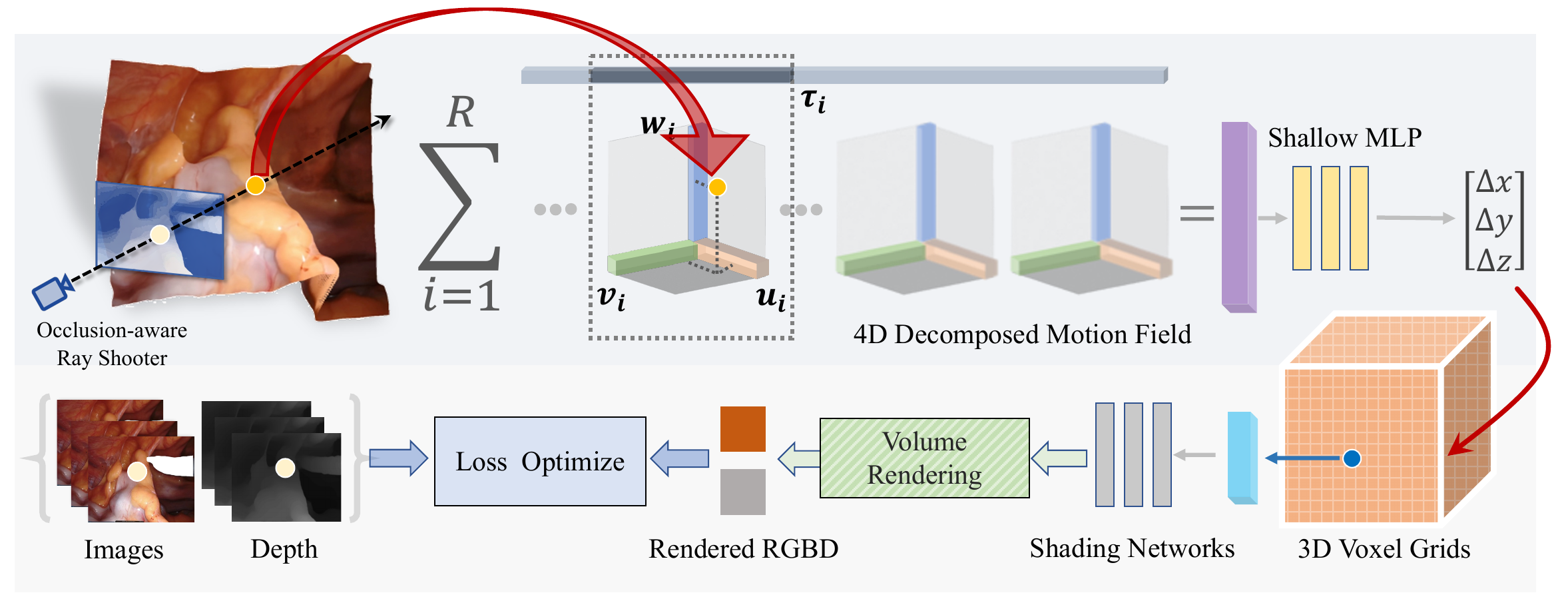}
    \caption{Pipeline of our proposed FastEndoNeRF framework, consisting of a 4D-decomposed motion field and dense 3D voxel grids.}
    \label{fig:fastendonerf_net_arch}
\end{figure}

We first aim to propose a dynamic scene representation to model soft tissue's 3D shapes and textures from a stereo video clip of a dynamic surgical scene. Then we devise a particular de-occlusion rendering and stereo depth-supervised loss for optimizing the scene representation. Finally, we fill the reconstructed mesh surfaces into closed meshes and perform soft-body simulations on the filled meshes. The detailed descriptions are as follows.

\subsection{Efficient EndoNeRF Scene Representations}

In order to enable high-fidelity reconstruction of the surgical simulation environments, we resort to neural radiance fields. The fundamental neural radiance fields \cite{mildenhall2020nerf} for 3D scene representations are modeled in a coordinate-based MLP. Optimizing such scene representation to convergence is slow.
Alternatively, we adopt an implicit-explicit voxel grids-based scene representation, which is shown to achieve much faster optimization \cite{muller2022instant,sun2021direct,yu2021plenoxels,fridovich2023k}. Specifically, we model the shape and appearance of the scene in density volume grids $\mathbf{V}_\sigma \in \mathbb{R}^{H\times W \times D}$ and feature volume grids $\mathbf{V}_a \in \mathbb{R}^{C\times H\times W \times D}$, where $H$, $W$, and $D$ are the resolutions for the x, y and z dimensions and $C$ is the channel number of the appearance features. For the density volume grids $\mathbf{V}_\sigma$, each grid vertex maintains its occupancy probability. For the feature volume grids $\mathbf{V}_a$, each grid vertex holds an appearance code. To map the appearance code into RGB color, we introduce a shallow MLP $S_{\Theta}: \mathbb{R}^C\rightarrow \mathbb{R}^3$ as a learnable implicit shading module. The geometry and appearance of any point $\boldsymbol{x}$ in the 3D space can be retrieved via tri-linear interpolation (denoted as $\operatorname{interp}(\cdot)$) of the 8 surrounding vertices' densities and features, i.e., the density $\sigma(\boldsymbol{x})=\operatorname{interp}(\boldsymbol{x}, \mathbf{V}_\sigma)$ and the color $\mathbf{c}(\boldsymbol{x})=S_{\Theta}(\operatorname{interp}(\boldsymbol{x}, \mathbf{V}_a))$.


Next, we consider surgical scene deformations. A dynamic surgical scene can be decomposed into a canonical radiance field and a time-dependent deformation field \cite{pumarola2021d,park2021nerfies}. Thereby the dynamic scene at time $t$ can be viewed as the canonical field warped by the deformation field at $t$. In our proposed method, the canonical radiance field is represented by the $\mathbf{V}_\sigma$ and $\mathbf{V}_a$. To support large and topology-varying deformations, we adopt decomposed 4D motion fields and a 3-layer MLP to model the deformation field, which maps a spatial-temporal coordinate $(\boldsymbol{x}, t)$ into its corresponding displacement $\Delta\boldsymbol{x}$.
In specific, we define a motion feature field as a $H\times W \times D \times T \times C_T$ tensor $\mathcal{T}$ \cite{chen2022tensorf}, where $T$ is the resolution of the time dimension and $C_T$ is the temporal feature channel number. Direct dense 5D modeling of the motion feature field is costly in storage and over-high-dimensional for optimization on sparsely captured frames. Thus, we need to seek another compact representation. Since deformations can be locally continuous and low-rank, as observed in \cite{cao2023hexplane,fridovich2023k}, we can decompose this tensor via outer product (Eq. \ref{eq:4d_deform_decomp}):
\begin{equation}
    \mathcal{T} = \sum_{r=1}^{R_1} \tau_r^4 \circ \mathcal{V}_r^{1,2,3} \circ b_r^4 + \sum_{r=1}^{R_2} \tau_r^3 \circ \mathcal{V}_r^{1,2,4} \circ b_r^3 + \sum_{r=1}^{R_3} \tau_r^2 \circ \mathcal{V}_r^{1,3,4} \circ b_r^2 + \sum_{r=1}^{R_4} \tau_r^1 \circ \mathcal{V}_r^{2,3,4} \circ b_r^1,
    \label{eq:4d_deform_decomp}
\end{equation}
where $R_1, R_2, R_3$ and $R_4$ are expected rank for each dimension, $\tau_r^l$ is a 1-D vector of the $l$-th dimension, $b_r^l$ is a feature basis of the $l$-th dimension, and $\mathcal{V}_r^{i,j,k}$ is a 3-D volume encompassing $i,j,k$-th dimensions. For each continuously queried point $(\boldsymbol{x}, t)$, we tri-linearly interpolate component tensors $\tau_r^l$ and $\mathcal{V}_r^{i,j,k}$ to obtain a motion feature vector. Then we feed the motion feature vector into a 3-layer MLP $G_\phi$ to compute the output displacement vector. In this way, the corresponding coordinates in the canonical field can be obtained by $\boldsymbol{x}' = \boldsymbol{x} + \Delta\boldsymbol{x}(\boldsymbol{x},t)$ with $\Delta\boldsymbol{x}(\boldsymbol{x},t) = G_\phi(\operatorname{interp}(\boldsymbol{x}, t, \mathcal{T}))$.



\subsection{Rendering and Optimization}

\textbf{Volume rendering.} With this scene representation, we can reconstruct the deformable surgical scene by optimizing the loss between rendered color $\hat{\mathbf{C}}$ and ground truth color $\mathbf{C}$. Specifically, the rendered color of the ray $\mathbf{r}(z)=\mathbf{o}+z\mathbf{d}$ at time $t$ can be evaluated by volume rendering as shown in Eq. \ref{eq:fastendonerf_vol_rendering}:
\begin{equation}
    \hat{\mathbf{C}}(\mathbf{r}(z), t) = \sum^M_{j=1} w_j \mathbf{c}_j, \quad
    w_j = \exp\bigg(-\sum_{i=1}^{j-1}\sigma_i \delta_i \bigg) \bigg(1-\exp\big(-\sigma_j \delta_j \big)\bigg),
    \label{eq:fastendonerf_vol_rendering}
\end{equation}
where $M$ is the number of sampled points along $\mathbf{r}(z)$, $\delta_i$ is the sampling step length, $\sigma_j$ and $\mathbf{c}_j$ are the density and color of the j-th sample evaluated by $\sigma(\boldsymbol{x}_j + \Delta\boldsymbol{x}(\boldsymbol{x}_j,t))$ and $\mathbf{c}(\boldsymbol{x}_j + \Delta\boldsymbol{x}(\boldsymbol{x}_j,t))$, respectively. The attenuation term $w_j$ can be regarded as the probability that the ray is transmitted to the j-th sample.


\vspace{0.3cm}\noindent\textbf{De-occlusion of surgical tools.}
According to the literature \cite{long2021dssr,wang2022neural}, soft tissues in surgical videos can often be occluded by surgical tools in the foreground. To address this issue and accurately reconstruct the soft tissues, our approach focuses on training the rays corresponding to tool pixels. Following the methodology proposed in EndoNeRF \cite{wang2022neural}, we generate binary tool masks for the left view of each frame. Instead of the mask-guided ray sampling proposed in EndoNeRF \cite{wang2022neural}, which bypasses rays per training iteration, we pre-compute all possible camera rays and check for intersections between these rays and the tool masks prior to training. This saves the computational costs during the scene optimization procedure, resulting in faster training. Any rays that pass through the tool masks are excluded from the training process. During training, the training batch $\mathcal{R}$ is randomly sampled from the pre-computed rays that have been screened in this manner. By doing so, we ensure that the optimization of the scene representation bypasses the tool pixels. Leveraging the auto-interpolation property of radiance fields, we can patch the occluded soft tissue areas using information from adjacent frames throughout the training procedures.

\vspace{0.3cm}\noindent\textbf{Distillation of stereo correspondence.} To exploit stereo geometry in confined in-vivo input, we propose to leverage stereo geometry to enrich 3D clues over the optimization of the scene representation. Very recent work unimatch \cite{xu2022unifying} learns dense correspondence on general vision datasets in a unified formulation for optical flow, stereo matching, and depth estimation tasks. Due to its superior performance over the previous method \cite{li2021revisiting}, we propose to distill stereo correspondence learned on general data into the surgical data along with the optimization of the surgical scene. To measure the learned stereo correspondence of the scene representation, we render depth from the radiance fields via $\hat{\mathbf{D}}(\mathbf{r}(z),t) = \sum^M_{j=1} w_j z_j$, where $z_j$ is the distance of the j-th sample along the ray $\mathbf{r}(z)$. The rendered depth is expected to converge to the estimated stereo depth once well-matched stereo correspondence is attained by optimizing the scene representation.
Thus, we estimate stereo depth $\mathbf{D}(\mathbf{r}(z),t)$ by stereo-matching the feature correspondence of the robotic surgery videos from unimatch \cite{xu2022unifying}. Lastly, we add a depth-supervised loss to the objective function, resulting in the final loss function:
\begin{equation}
  \begin{split}
    \mathcal{L} &= \sum_{\mathbf{r}(z) \in \mathcal{R}, t\in[0,1]} \left \| \hat{\mathbf{C}}(\mathbf{r}(z), t) - \mathbf{C}(\mathbf{r}(z), t) \right \|_2^2 + \lambda_d \operatorname{Huber}\left( \hat{\mathbf{D}}(\mathbf{r}(z), t), \mathbf{D}(\mathbf{r}(z), t) \right),
  \end{split}
  \label{eq:fastendonerf_obj_func}
\end{equation}
where $\mathbf{C}(\mathbf{r}(z),t)$ and $\mathbf{D}(\mathbf{r}(z),t)$ is the corresponding ground truth pixel color and unimatch \cite{xu2022unifying} stereo depth of camera ray $\mathbf{r}(z)$ at the time $t$. Here we adopt Huber loss \cite{huber1992robust} which is more stable to outliers. Compared with the stereo depth maps predicted by STTR \cite{li2021revisiting}, supervising better depth maps via unimatch can further decrease the training time since the depth refinement module proposed in EndoNeRF \cite{wang2022neural}, which requires depth rendering of all training images, is no longer needed.

\begin{algorithm}
	\caption{mesh-open2closed (Sec. \ref{sec:extract_closedmesh})}\label{alg:close_mesh}
	\begin{minipage}{0.42\textwidth}
	\KwData {Open mesh surface $\mathcal{M} = (\mathcal{V}, \mathcal{F})$ \cg{which has one hole}, soft tissue thickness $\zeta$}
	\KwResult {Closed mesh surface with updated $\mathcal{V}$ and $\mathcal{F}$}
	$\mathcal{A} \gets \text{non-manifold edges of $\mathcal{F}$}$\;
				\For{each vertex $v \in \mathcal{V}$}{
							$adj_v \gets \{\}$\;
						}
				\For{each edge $\{u,v\}\in \mathcal{A}$}
				{$adj_u  \gets adj_u  \cup \{v\}$\;
				$adj_v  \gets adj_v  \cup \{u\}$\; }
				\tcc{Conduct BSF to sort boundary vertices into an ordered list $\mathcal{C}$}
				$\{u,v\} \gets \operatorname{Head(\mathcal{A})}$ \;
				$\mathcal{C} \gets \{u\}$ \;
				\While{$v \neq u$}
			{	$\operatorname{Enqueue(\mathcal{C},v)}$ \;
				$u \gets v$\;
				\For{each $k\in adj_v$}
				{\If{$k \notin \mathcal{C}$}
			{	$v \gets k$\;
				break\;}
				} }
				\end{minipage}
				\hfill\vline\hfill
				\begin{minipage}{0.475\textwidth}
					$\operatorname{Enqueue(\mathcal{C},\operatorname{Head(\mathcal{C})})}$\;
				 $center \gets (\frac{1}{\left|\mathcal{V}\right|}\sum_{v\in \mathcal{V}} X_v, \frac{1}{\left|\mathcal{V}\right|}\sum_{v\in \mathcal{V}} Y_v, \zeta)$\;
				 $\mathcal{V} \gets \mathcal{V} \cup \{center\}$\;
				 $first \gets \operatorname{True}$\;
                  \tcc{Iteratively build a base plane of the soft tissue surface and connect the base with the boundary vertices along $\mathcal{C}$}
				 \While{$!\operatorname{isEmpty(\mathcal{C})}$} {
				 	$v \gets \operatorname{Dequeue(\mathcal{C})}$\;
				 	\If {$!first$} {
				 		\tcc{Projection of the last $v$}
				 		$p \gets v'$\;}
				 		\tcc{Projection of $v$ in the appended base of the soft tissue}
				 		$v' \gets (X_v, Y_v, \zeta)$\;
				 		$\mathcal{V} \gets \mathcal{V} \cup \{v'\}$\;
				 		\If {$!\operatorname{isEmpty(\mathcal{C})}$} {
				 			$u \gets \operatorname{Head(\mathcal{C})}$\;
				 			$\mathcal{F} \gets \mathcal{F} \cup \{(v, u, v')\}$\;
				 		}
				 		\If {$!first$} {
				 			$\mathcal{F} \gets \mathcal{F} \cup \{(p, v, v')\}$\;
				 			$\mathcal{F} \gets \mathcal{F} \cup \{(p, center, v')\}$\;
				 		}
				 		$first \gets \operatorname{False}$\;
				 	}
				\end{minipage}
\end{algorithm}

\subsection{Extraction of Closed Meshes for Soft-Body Simulations}
\label{sec:extract_closedmesh}

\begin{figure}[t!]
    \centering
    \includegraphics[width=\textwidth]{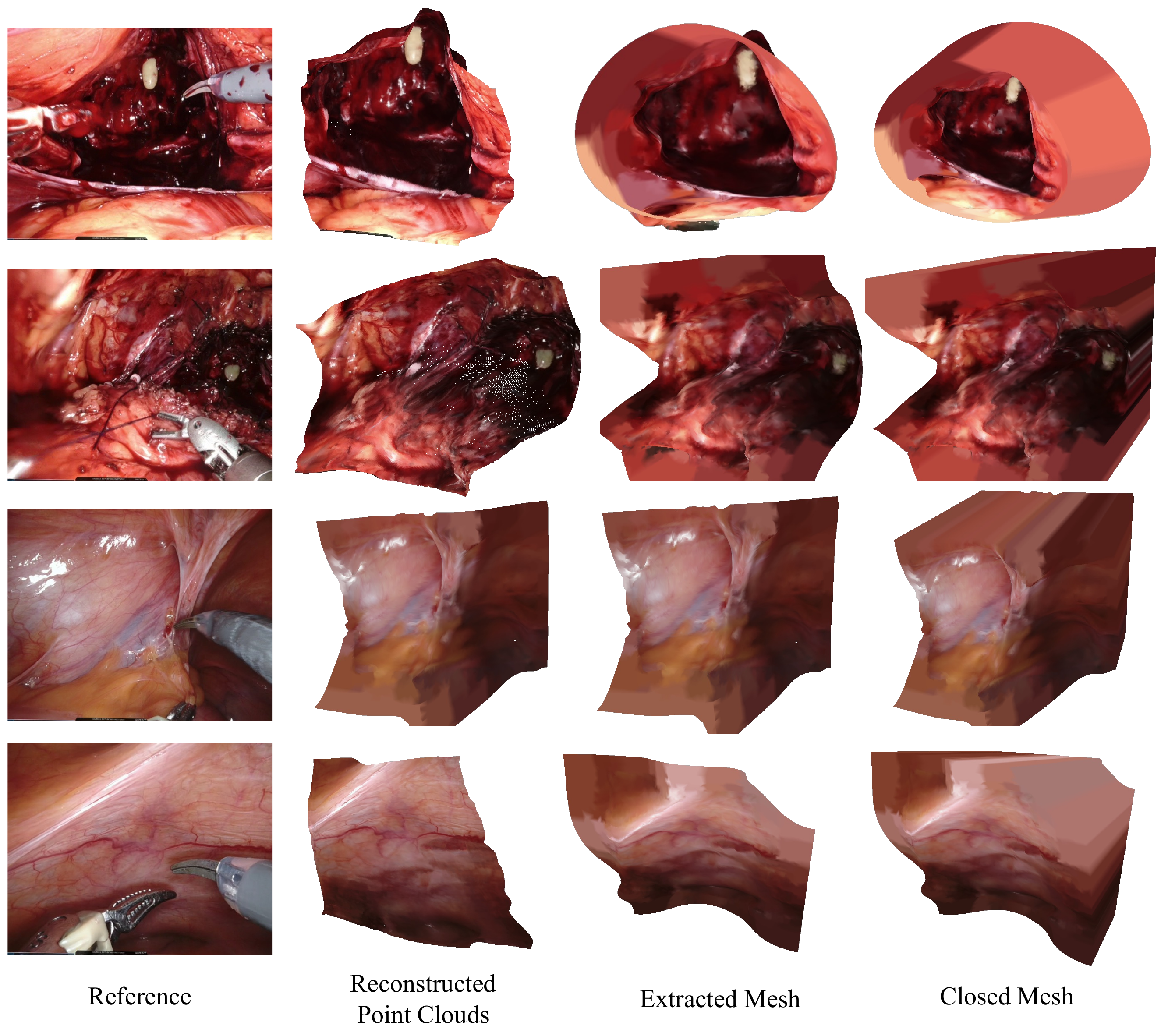}
    \caption{Reconsruction results. The first column gives the reference input image, the second column exhibits reconstructed point clouds, the third column shows the meshes obtained by Poisson surface reconstruction on the point clouds, and the last column displays the closed meshes appended with a base.}
    \label{fig:fastendonerf_recon_results}
\end{figure}

After we obtain an optimized dynamic radiance field, we aim to perform physically-based simulations on the reconstruction.
Numerically solving physically-based simulation systems requires dividing the object material domain into a number of geometry primitives. Since our reconstructed scene representation only encodes the seen soft tissue surface in an implicit geometry, we need to first obtain its explicit form and convert it into a simulatable object. To do this, we propose the following procedure. We first render the reconstructed canonical radiance fields to color and depth maps. Then, we back-project RGB-D maps into point clouds. Namely, each 3D point $(x, y, z)$ can be computed from a corresponding pixel $(P_u, P_v)$ with depth value $\hat{\mathbf{D}}$ as $(x, y, z)=(\hat{\mathbf{D}}(P_u - C_x) / f, \hat{\mathbf{D}}(P_v - C_y) / f, \hat{\mathbf{D}})$, where $(C_x, C_y)$ is the principal point and $f$ is the focal length. Bilateral filtering is also applied to smooth the point clouds.
After conversion to point clouds, we perform Poisson surface reconstruction to extract the mesh surface from the simplified point clouds.
Subsequently, we need to construct supporting structures underneath the surface for deformable object simulations.
Material Point Method (MPM) and Finite Element Method (FEM) both require a closed mesh surface as the input for discretization.
For the MPM solvers, dense particles are sampled to fill the soft tissue surface \cite{wang2020massively}. As for the FEM solver, robust tetrahedral meshing algorithms \cite{hu2018tetrahedral,hu2020fast,hang2015tetgen} are proposed to convert surface objects into tetrahedrons.
Thus, we tailor an efficient \textit{mesh-open2closed} algorithm that can universally enclose the reconstructed mesh surfaces. The pseudocode of the algorithm is given in Algorithm \ref{alg:close_mesh}, where the input mesh vertices $\mathcal{V}$ and triangles $\mathcal{F}$ are structured in 2D arrays, and $X_v$, $Y_v$ denote the x and y dimensions of vertex $v$. The algorithm begins with constructing the boundary edges of the reconstructed surface and organizing them in a list. Those boundary edges can be classified by non-manifold test, i.e., manifold edges should be simultaneously included in 2 triangles. After finding non-manifold edges, we can conduct a breadth-first search (BFS) to sort boundary vertices into an ordered list. Then, we iteratively build a base plane of the soft tissue surface in the shape of the open mesh boundary and connect the base with the boundary vertices along the ordered list. During this procedure, we loop for each vertex $v$ in the ordered edge list and find the projection of $v$ in the appended base of the soft tissue. Then we connect the projection of the last $v$, the projection of the current $v$, and the center of the base plane to create new faces.
\cg{It is noteworthy that our algorithm is designed for the input mesh with a single ``hole'', i.e., there is only one connected edge. This assumption usually holds since the incisions on soft tissues are relatively shallow in in-vivo surgical scenes. If there are two disjoint surfaces represented in the reconstructed field, a solution is to run the algorithm separately for each surface.}

\begin{table}[t!]
    \centering
    \caption{Quantitative evaluation and comparison of our method and baselines. We evaluate photometric errors and training time of the dynamic reconstruction.}
    \begin{tabular}{ >{\raggedright\arraybackslash}m {6.8em} | >{\centering\arraybackslash}m {8em} | >{\centering\arraybackslash}m {8em} |  >{\centering\arraybackslash}m {8em} |  >{\centering\arraybackslash}m {3.5em} }
        \toprule[0.5pt]
        \textbf{Methods} & \textbf{PSNR} $\uparrow$ & \textbf{SSIM} $\uparrow$ & \textbf{LPIPS} $\downarrow$ & \textbf{Time} $\downarrow$ \\ \hline
        E-DSSR~\cite{long2021dssr} & \cellcolor{red!40}13.398 $\pm$ 1.387 & \cellcolor{red!40}0.630 $\pm$ 0.057 & \cellcolor{red!40}0.423 $\pm$ 0.047 & \cellcolor{gray!40}\cm{0} \\
        EndoNeRF~\cite{wang2022neural} & \cellcolor{green!70}29.831 $\pm$ 2.208 & \cellcolor{green!70}0.925 $\pm$ 0.020 & \cellcolor{green!70}0.081 $\pm$ 0.022 & \cellcolor{red!60}> 10h \\ \hline
        FastEndoNeRF & \cellcolor{green!70}29.586 $\pm$ 3.48 & \cellcolor{green!25}0.892 $\pm$ 0.040 & \cellcolor{green!25}0.145 $\pm$ 0.050 & \cellcolor{green!70}27min \\ \bottomrule[0.5pt]
    \end{tabular}
    \label{tab:fastendonerf_quanti_eval}
\end{table}

\section{Experiments}

\subsection{Evaluation of Efficient EndoNeRF}

\begin{figure}[t!]
    \centering
    \includegraphics[width=\textwidth]{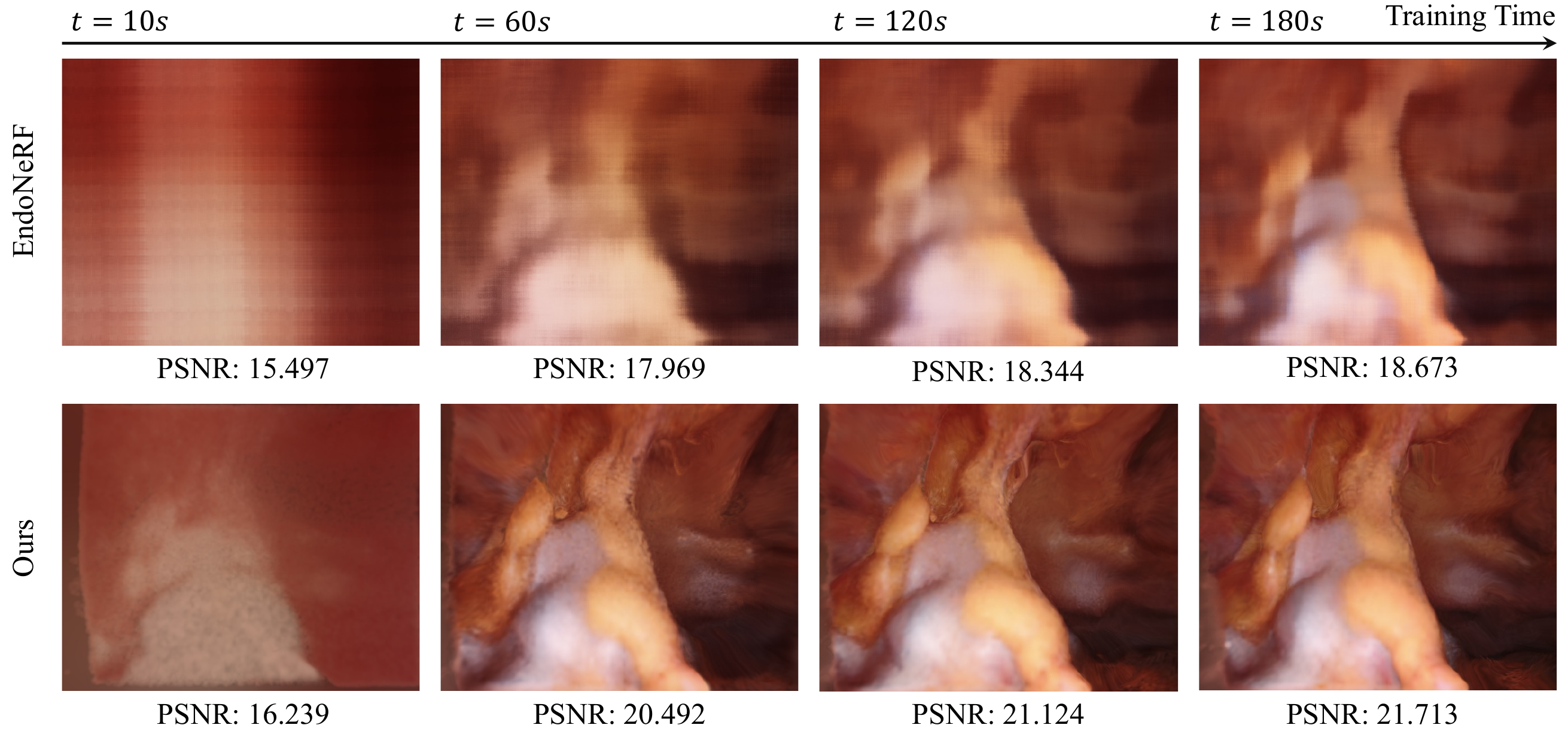}
    \caption{Comparisons of reconstruction quality between EndoNeRF and our method within the first 3 minutes of training.}
    \label{fig:fastendonerf_time_cmp}
\end{figure}

We conducted an evaluation of our proposed method on a set of typical clips of robotic surgery videos, captured from 10 cases of our in-house DaVinci robotic prostatectomy dataset. In addition to the cases used in EndoNeRF \cite{wang2022neural}, the new cases contain suturing, bleeding, and cutting on soft tissues. Each clip lasted for 4 to 8 seconds and was sampled into 45 $\sim$ 180 frames. These clips were captured from stereo cameras, and they encompassed challenging scenes with non-rigid deformation and tool occlusion.
To establish the effectiveness of our new method, we compared it with two strong baselines: the recent NeRF-based method EndoNeRF \cite{wang2022neural} and the traditional DynamicFusion-based approach E-DSSR \cite{long2021dssr}. For qualitative evaluation, we exhibit the reconstruction objects produced by our method, including reconstructed point clouds, surface meshes, and closed meshes.
Due to clinical regulations, it is infeasible to collect ground truth depth for numerical evaluation on 3D structures. To perform quantitative comparisons, we instead used photometric errors, such as PSNR, SSIM, LPIPS, and training time, as evaluation metrics. This evaluation methodology is consistent with that used in previous work on surgical scene reconstruction, such as \cite{long2021dssr,wang2022neural}, and is widely used in the field of neural rendering.

Figure \ref{fig:fastendonerf_recon_results} showcases our reconstruction outcomes, including extracted point clouds, soft tissue mesh surfaces, and closed meshes.
Our FastEndoNeRF algorithm excels at reconstructing watertight surfaces of soft tissues from videos, faithfully capturing the intricate in-vivo textures. Despite the presence of large deformations, our method tracks the dynamics of the soft tissues using our proposed 4D decomposed motion field. For tool occlusion in the input videos, our method manages to patch tool-occluded areas by leveraging information from adjacent frames, ensuring a comprehensive and watertight representation of the dynamic soft tissue.
In order to ensure that the reconstructed surface is suitable for simulation purposes in contemporary simulation engines, we have employed a mesh extraction scheme capable of constructing high-resolution meshes with intricate textures and shapes from the reconstructed point clouds. Furthermore, our proposed \textit{mesh-open2closed} algorithm facilitates the creation of a closed structure by appending a base to the mesh surface. This closed structure is essential for enabling accurate simulations in the chosen environment.
In Figure \ref{fig:fastendonerf_time_cmp}, we run our method and the original EndoNeRF \cite{wang2022neural} on the same NVIDIA RTX 3090 GPU for 3 minutes and compare their training efficiency.
\cm{Due to} the limited training time, the reconstruction results obtained with EndoNeRF remain noisy and blurry. Conversely, our method demonstrates impressive performance even at an early training stage (i.e., 10s to 60s), with the ability to approximate the scene's appearance and shape accurately. This validates the superior training convergence speed of our proposed scene representation. \cm{It is noteworthy that our model employs $\sim$160M parameters, consuming 4GB GPU memory for training each case. Without factorizing the 4D deformation field, the 4D deformation field would necessitate an allocation of over 12GB of memory during the training procedure, which shows the effectiveness of our compact dynamic scene representations.}

\begin{figure}[t!]
    \centering
    \includegraphics[width=\textwidth]{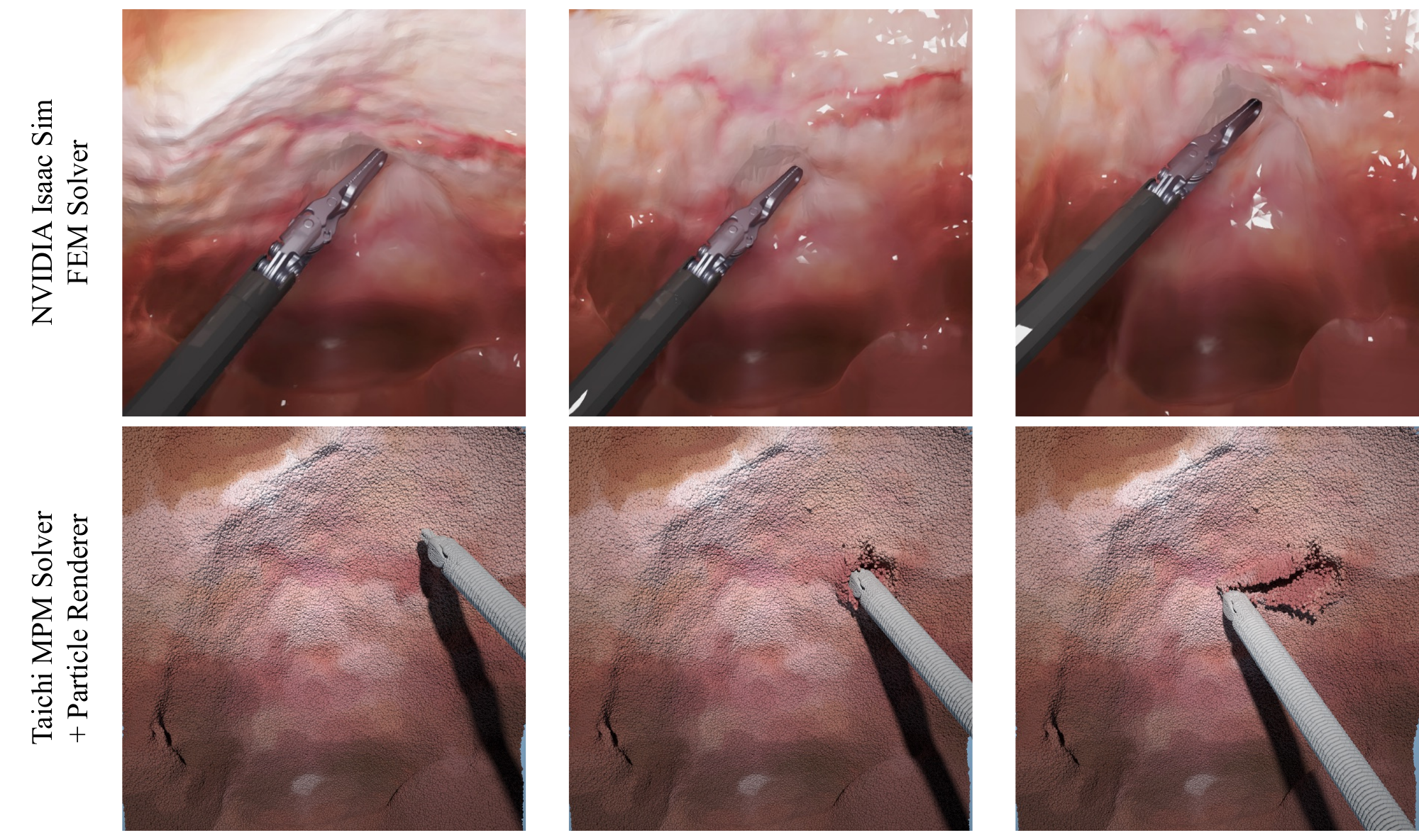}
    \caption{Soft tissue simulation results. The first row exhibits real-time interaction between surgical tools and reconstructed soft tissues in NVIDIA Isaac Sim \cite{liang2018gpu}. The second presents a simulation example of soft tissue incision with MLS-MPM algorithm \cite{hu2018moving} implemented in Taichi \cite{hu2019taichi}.}
    \label{fig:app_surgical_sim}
\end{figure}

Table \ref{tab:fastendonerf_quanti_eval} displays a \cm{quantitative} comparison of the metrics PSNR, SSIM, LPIPS, and training time. Both methods exhibit impressive photometric results when compared to the traditional method of E-DSSR \cite{long2021dssr}. Despite a slight decrease in performance, FastEndoNeRF achieves a remarkable training time improvement of approximately 20 times faster than EndoNeRF. By training FastEndoNeRF for 27 minutes, we can achieve comparable quality to EndoNeRF trained for over 10 hours. This highlights the efficiency and effectiveness of the FastEndoNeRF approach.



\subsection{Initial Application for Surgical Scene Simulation}
\label{sec:fastendonerf_application}

Virtual surgical training platforms have become increasingly significant in surgery education and training \cite{long2022robotic,long2023human}. However, building a surgical education and training platform is associated with several challenges, including limited exposure to real-life surgical cases, and limited access to high-fidelity simulation. Our proposed framework can overcome these challenges by providing a reconstructed realistic environment for surgical trainees to practice and master their skills.

\vspace{0.3cm}\noindent\textbf{Real-time FEM simulation.} Here we first build a real-time virtual surgery simulation in NVIDIA Isaac Sim \cite{liang2018gpu}, where FEM is the solver for simulating reconstructed continuum objects. In the first row of Figure \ref{fig:app_surgical_sim}, we import a reconstructed closed mesh into NVIDIA Isaac Sim and tune its physical properties to make it behave like soft tissues. Owing to advanced GPU acceleration, NVIDIA Isaac Sim enables real-time FEM simulation and rendering, producing high-fidelity deformations under the dissection interaction.
The automatic reconstruction of the simulation environment from real surgical videos ensures that the in-vivo textures are accurately preserved, thereby enhancing the visual realism of surgical simulations. \cm{The proposed algorithm for closed mesh extraction \cg{facilitates} material domain discretization for the FEM solver within Isaac Sim. If the imported meshes are not closed in Isaac Sim, the mesh tetrahedralization procedure will fail, resulting in unreasonable simulation effects.} Moreover, the creation procedure for this simulation environment is highly scalable, thanks to the efficiency of the surgical reconstruction pipeline.



\vspace{0.3cm}\noindent\textbf{MPM simulation.} While the FEM solver in NVIDIA Isaac Sim achieves basic soft-body simulation, it lacks the ability to perform damage operations on continuum objects, which is considered a crucial aspect of simulating soft tissues. In order to address this limitation, we employ the Material Point Method (MPM) \cite{sulsky1995application}, a hybrid grid-particle method that combines the strengths of both Eulerian and Lagrangian approaches. This method enables us to handle large deformations and complex material behavior, as demonstrated in recent papers \cite{wolper2019cd,wolper2020anisompm}. To specifically support damage deformations on soft bodies and achieve two-way coupling between rigid and non-rigid objects, we implement the state-of-the-art MLS-MPM \cite{hu2018moving}. In Figure \ref{fig:app_surgical_sim}, the second row illustrates an example of soft tissue damage resulting from dissection. It is evident that MLS-MPM is capable of accurately capturing the incision behavior on the soft tissues. While MPM offers soft tissue damaging simulation, it is characterized by high computational costs and falls short of achieving real-time simulations. \cm{In the simulation stage, $\sim$5M particles are generated for simulation, resulting in a memory consumption of around 5GB.}



\section{Conclusion}


We present an innovative and data-driven framework for constructing surgical simulation environments using endoscopic videos. Our approach introduces a new fast dynamic scene representation based on NeRF, which significantly accelerates the 3D reconstruction process of surgical scenes. Additionally, we propose a closed mesh extraction algorithm that converts reconstructed soft tissue surfaces into simulation objects. To demonstrate the versatility and applicability of our framework, we showcase multiple simulations of reconstructed surgical environments for diverse clinical applications. Our proposed methodology aims to inspire a significant advancement in the field of surgical simulation and is poised to open up new possibilities for next-generation surgical training and surgical robot learning.

\vspace{0.3cm}\noindent\cm{\textbf{Limitations and future work.} There are still some under-explored problems with our current methods. First, the de-occlusion of surgical tools relies on the interpolation of radiance fields, which will cause artifacts in the textures of occluded soft tissues. This could be solved by incorporating generative models to inpaint the textures. Second, as an initial trial, our simulation is based on the naive versions of FEM and MPM. In the future, we aim to test more simulation algorithms on our reconstructed soft tissues, e.g., XFEM and XMPM, to achieve more realistic simulation effects.}

\vspace{0.3cm}

\backmatter





\bmhead{Acknowledgments}

This work was supported in part by a grant from the Research Grants Council of the Hong Kong Special Administrative
Region, China (Project No. 24209223), in part by the Hong Kong Innovation and Technology Fund (Project No. ITS/223/22), in part by the Science, Technology and Innovation Commission of Shenzhen Municipality Project No. SGDX20220530111201008, and in part by InnoHK Multi-Scale Medical Robotics Center.

\bibliography{references}

\end{document}